\documentclass[sigplan,screen]{acmart}
\makeatletter
\def\@ACM@checkaffil{
    \if@ACM@instpresent\else
    \ClassWarningNoLine{\@classname}{No institution present for an affiliation}%
    \fi
    \if@ACM@citypresent\else
    \ClassWarningNoLine{\@classname}{No city present for an affiliation}%
    \fi
    \if@ACM@countrypresent\else
        \ClassWarningNoLine{\@classname}{No country present for an affiliation}%
    \fi
}
\makeatother
\usepackage{graphicx} 
\usepackage{caption}
\usepackage{subcaption}
\usepackage{pgfplots}

\title{The Impact of Virtual Achievements on Online Learning Applications}
\author{Venkata Sai Bhargav Bathini}
\affiliation{%
  \institution{University of Florida}
}
\email{vbathini@ufl.edu}
\author{Lokesh Meesala}
\affiliation{%
  \institution{University of Florida}}
\email{lokeshmeesala@ufl.edu}
\author{Ahamad Shaik}
\affiliation{%
  \institution{University of Florida}}
\email{ahamadshaik@ufl.edu}
\author{Hrushyang Adloori}
\affiliation{%
  \institution{University of Florida}}
\email{adloorih@ufl.edu}

\renewcommand\footnotetextcopyrightpermission[1]{}

\begin{document}

\pagestyle{plain}

\renewcommand{\shortauthors}{Bhargav, Lokesh, Ahamad, Hrushyang.}
\begin{abstract}
In recent times a number of platforms are using badge-based achievements or leaderboards to increase user involvement and participation. Although there are some positive informal reports it still needs evidence to support the claims and efficiency of the virtual achievements. Due to recent advancements, there is a question of up to what extent virtual achievement systems have on users using particular platforms.

Here in this paper, we discuss measuring the impact of the leaderboard-based achievement system by integrating it into an online learning android application UPSC Pre that has thousands of questions and answers categorized topic wise related to UPSC exams which are one of the toughest exams to crack in the world. We are conducting the experiment on 10 randomly chosen students who are using the app in a controlled setting and the data measurement is done using the Firebase Analytics tool by Google. We observed that the students using the leaderboard have increased participation without any reductions in their quality and the time of using the platform has increased compared to previous engagements. Students who participated in the experiment felt the leaderboard was competitive and enjoyed gaining positions on the leaderboard and wanted it in the User interface for other platforms as well.
\end{abstract}

\maketitle

\section{Introduction}
 The First implementation of achievements which are known as badges in leaderboard was implemented for games in 2005 where achievements are added to then 3 year old Xbox Live platform~\cite{1}. From then the usage of badges has spread to other gaming platforms which prove to be successful and has led to increased sales revenues. Due to its potential, the badge system was further transferred to platforms in other domains.
 
 Virtual achievements, such as badges, points, or leaderboards, have become increasingly popular in educational settings as a way to incentivize and reward students for their learning efforts. These achievements serve as a recognition of students' progress and can enhance their motivation and interest to engage with the material  ~\cite{2}. Despite the growth and usage, it still needs evidence to support the claims and efficiency of the Virtual achievement systems. With the increase in the technology and the continuous development to play a crucial role in learning environment, to understand how virtual achievements can impact student engagement and there is a need to research and present efficient results in this area.
 
 The purpose of our research is to prove the hypothesis, 'the use of virtual achievements through introducing leaderboards in a multiple topic learning App with no prior leaderboard system helps in improving student engagement in educational settings' ~\cite{3}. Specifically, the study seeks to answer the question: Do establishing leaderboards enhance student engagement and motivation to learn? We plan to prove the hypothesis by introducing a leaderboard system in an existing application and study the subject performance.
 
 Through a comprehensive review of the literature and empirical data collection, this study will examine the impact of virtual achievements on student engagement, that includes introducing a leader-board to the participants and how different students respond to these incentives. We study the subject during a period of 2 months with each month dedicated to a test case i.e., one month without leaderboard and one month. The independent variables in the study are the questions in the application and the introduced leaderboard system. The levels for the independent variable, leaderboard involves two cases of with no leaderboard and with leaderboard. The dependent variable in the study is the user engagement time in the app.
 
 The results of both cases are compared after the experiment to see a significant increase in the user engagement after introducing the virtual achievement system. The results of this study can inform educational technology designers and educators on how to design the platforms to increase the interaction of users and also to improve the quality and effectiveness of educational tools and techniques for enhancing student engagement and learning outcomes. The results support the positive impact of inducing leaderboards i.e. yielding an increase in average user engagement time with the application. 

 \section{Background}
The research paper on the virtual achievements effect on engagement of students by Kuo et al. investigates the impact of virtual achievements in an online learning environment ~\cite{4}. A quasi-experimental study with two groups of undergraduate students taking an online course in psychology where One group was given access to virtual achievements in the form of badges and points, while other group did not have access to any virtual achievements. The authors collected data on student engagement through log data, surveys and found that students who had access to virtual achievements showed higher levels of engagement, including higher levels of participation and motivation.
 
 The impact of digital badges on student engagement study conducted by Khoury et al ~\cite{5} suggests digital badges will have positive impact on student engagement. The study involved 78 undergraduate students taking an online course, with half of the students randomly assigned to earn digital badges for completing course activities. Data was collected through surveys and course activity tracking. The study found that students who earned digital badges were more engaged than those who did not earn badges. 

 A study conducted on introducing leaderboards to improve student motivation by Hsu et al closely relates to our experiment ~\cite{6}. The use of a leaderboard in a computer programming course will increase student motivation and learning outcomes. The study involved two groups of students - one group with a leaderboard and one group without and the group with the leaderboard had higher motivation and learning outcomes compared to the group without. 

 Gamification and leaderboard performance in higher education by Villagrasa et al ~\cite{7}. The study involved a group of students who were enrolled in a gamified course with a leaderboard. Data was collected through surveys and analysis of student performance. The leaderboard was found to increase student engagement and motivation, but did not have a significant impact on learning outcomes.

 The research on impact of reward on student engagement by Manca and Ranieri involved 50 undergraduate students taking a flipped classroom course, with half of the students randomly assigned to receive rewards for completing course activities~\cite{8}. Data was collected through surveys and course activity tracking. The study found that students who received rewards were more likely to engage with course materials.

 A study conducted by Wang, Chen, and Liang sets to explore the impact of the same ~\cite{9}. The authors conducted a meta-analysis of previous studies and synthesized the findings on the impact of mobile learning on students' learning behaviors and performance. The authors found that mobile learning can improve students' learning behaviors and performance, particularly in terms of knowledge acquisition, critical thinking, and problem-solving skills.

\section{Experiment setup}
 We begin by describing the UPSC Pre app environment and an overview of the leaderboard system used in the study. UPSC Pre is an android app and consists of previous years questions of multiple exams like the Civil Services Examination(CSE), National Defence Academy(NDA), Combined Defense Services(CDS), Combined Armed Police Forces(CAPF) conducted by UPSC - Union Public Service Commission of India every year. This app contains multiple choice questions(MCQs) of all these exams categorized into different topics like History, Economics, Polity, Geography, and others. Each question has 4 options only one of them is correct. Each year approximately a million students appear for the exam and they students need to solve the previous year's questions multiple times to clear the exams because 5-6 questions of the exams are repeated from the previous years questions every year and each question is extremely important.
 
 The study is done with in the subjects. We chose a total of 10 students randomly who are already using the app and are familiar with the interface of the app. To ensure the balance of gender we choose 5 male and 5 female students for this study. The independent variables in the study are the questions in the application and the leaderboard system introduced. The levels for the independent variable, leaderboard involves two cases of with no leaderboard and with leaderboard. The dependent variable in the study is the user engagement time in the app.

 We chose the Leaderboard as our main motivating factor for students to increase their app engagement time in the app. The reason is that it increases the competitive spirit among the students and they want to increase their ranking in the leaderboard by solving more questions.  Further, for the purpose of the study, each student was informed about the study that is being conducted. An e-consent was designed that pops up for the participant to accept before they get involved with the experiment. There is also designated drop out option always included in the app if any of participant wants to opt out of the study at any time during the study. We have used engagement time as a metric to measure the motivation of the students in the app because it shows that the students have answered more questions in the app. The Figure 1 shows the screenshots taken from the app which is the test condition. The Fig.1(a) shows the various topics in the app. The Fig.1(b) shows the leaderboard that the participants see which is updated on a real-time basis.
 
We don't need any type of balancing technique because there are a thousands of questions to be chosen from where each question carries the same points for a correct answer and the randomness is justified. The app is already live on the playstore so we tested without leaderboard first and with leaderboard by updating the app after a month.

\begin{figure}
\centering
\begin{subfigure}{.25\textwidth}
  \centering
  \includegraphics[width=0.9\textwidth]{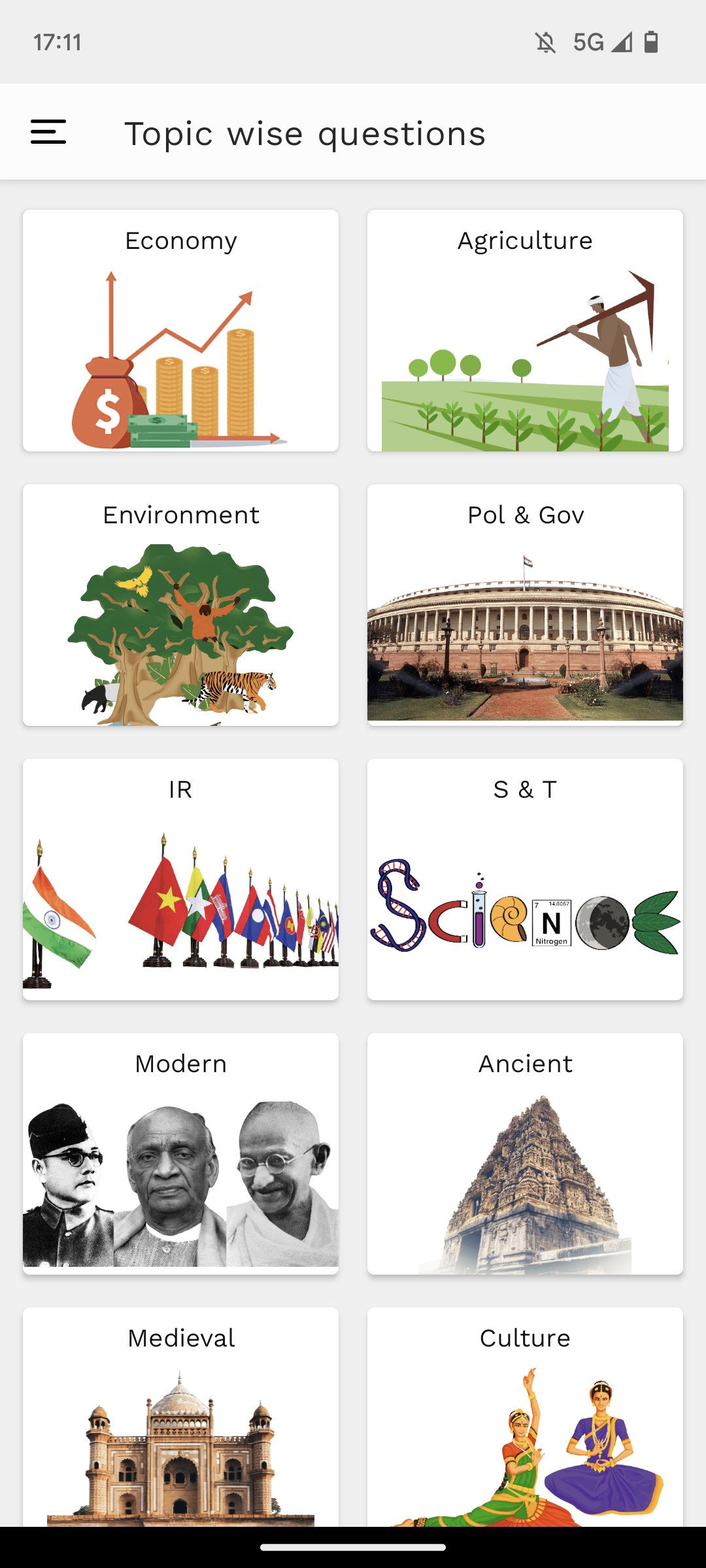}
  \caption{Topics}
  \label{fig:sub1}
\end{subfigure}%
\begin{subfigure}{.25\textwidth}
  \centering
  \includegraphics[width=0.9\textwidth]{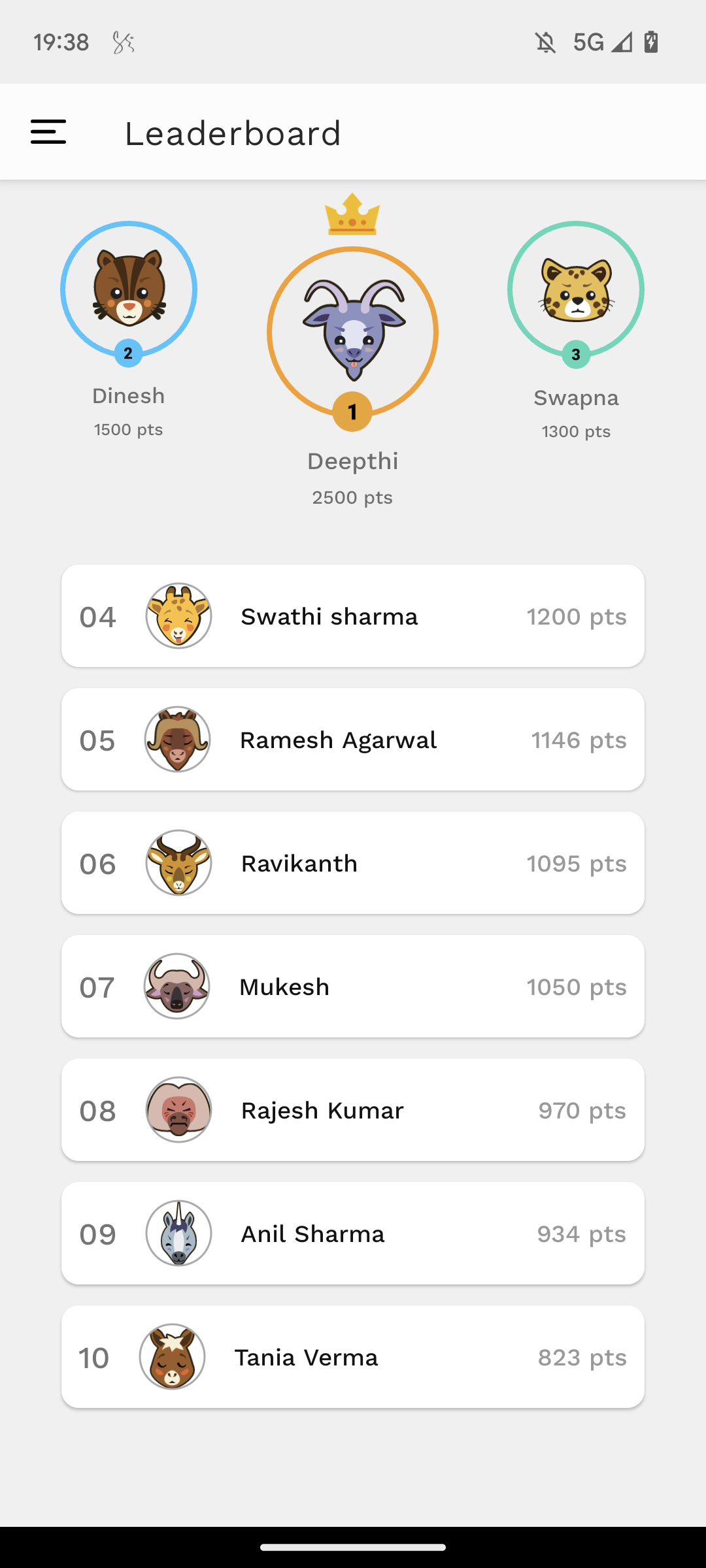}
  \caption{Leaderboard}
  \label{fig:sub2}
\end{subfigure}%
\caption{Screenshots from the UPSC Pre app - testing}
\label{fig:test}
\end{figure}

\subsection{Data Collection}
In order to use the app, the participants need an Android phone with Android version greater than 5.0 and up. The participants also need an understanding of different topics of the UPSC exams to use the app. We initially tested with analytic tools like Flurry Analytics, Mixpanel and Firebase Analytics by Google. After observations, finally Firebase Analytics is used in the app to collect the data because of the advanced features in it. The Firebase analytics measures the various events and of the specific users we are studying as a part of the study. 

A student is considered to be engaged in the app on a particular day if they either answer at least one question in the app be it correct or incorrect. The engaged time is the time spent in the app from the opening of the app to the closing of the app. The data of the 10 participants the study is transmitted daily to the Firebase servers. The Figure.1 shows the app that is used by the participants and the Figure.4 shows the apparatus that is the Firebase analytics used in the experiment. 

Participants can choose to answer questions from any of the 20 topics that are in the app. No data that was excluded from subsequent analysis because the data collected by the Firebase analytics is accurate. The leaderboard is realtime and updates whenever the users answers the questions.

There was no issues in the data collection as we have integrated the firebase third-party Analytics API by Google which is accurate in the measurement of the data.

\section{Data Analysis}

The dependent variable for the study is the user engagement time calculated for each test subject. First the user engagement for each subject is noted every day and sum of total engagement time is used as a measure for that week. In this pattern, for the month of February, we take 4 values for 4 weeks as the output for this case and calculate the mean and standard deviation for that subject. These values are tabulated for all the subjects under the column leaderboard off. the data is pre-processed and reorganized for the cases such as, if no activity is done by the user for a week then it is noted as 0 for that week. In case of 5 week months or less that 4 week months, the days are added or subtracted into the last month to remove the outliers and reorganise the data. Similarly, the resultant user engagement time is noted after implementing the leaderboard for the month of march and the results are tabulated.
\begin{table}[htp]
    \centering	
    \small
    \begin{tabular}{|l|l|l|}
    \hline
    \textbf{Test Subjects} & \textbf{Leaderboard off} & \textbf{Leaderboard On}\\\hline
    1 & mean = 22.6 & mean = 41.19 \\ & SD = 2.15 & SD = 4.23 \\\hline
    2 & mean = 26.4 & mean = 38.62 \\ & SD = 9.99 & SD = 5.63 \\\hline
    3 & mean = 24.2 & mean = 43.26 \\ & SD = 7.28 & SD = 6.21 \\\hline
    4 & mean = 23.8 & mean = 40.54 \\ & SD = 4.82 & SD = 5.42 \\\hline
    5 & mean = 28.1 & mean = 39.04 \\ & SD = 8.76 & SD = 2.15 \\\hline
    6 & mean = 30.0 & mean = 41.33 \\ & SD = 5.63 & SD = 2.77 \\\hline
    7 & mean = 29.0 & mean = 42.51 \\ & SD = 9.51 & SD = 4.82 \\\hline
    8 & mean = 25.7 & mean = 40.09 \\ & SD = 2.77 & SD = 3.47 \\\hline
    9 & mean = 25.1 & mean = 39.21 \\ & SD = 3.47 & SD = 3.64 \\\hline
    10 & mean = 27.3 & mean = 42.08 \\ & SD = 6.89 & SD = 6.89 \\\hline
    Total & Mean = 24.35 & Mean = 39.58 \\ & SD = 3.08 & SD = 1.85\\\hline 
    \end{tabular}
    \caption{Weekly engagement mean and SD table.}
    \label{tbl:pr1data1}
\end{table}
In Table 1, we recorded the average engagement time of each test subject for the month of February where leaderboard was not available to users. For the month of March, the 10 test subjects were informed of the leaderboard and were given access to check their ranking in the board. Engagement time is recorded every week and cumulative mean and standard deviation are tabulated for every subject. Figure 2 represents the graphical representation for the tabulated data for the two test conditions given. We calculated the mean and standard deviation for the tabulated data and the same is recorded as a bar graph.
\begin{figure}[htp]
    \centering
    \includegraphics[width=0.5\textwidth]{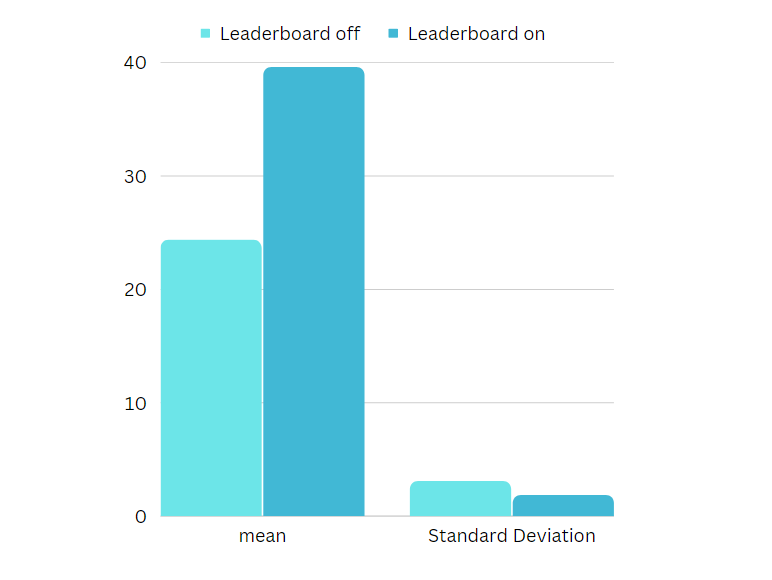}
    \caption{Graph plot of mean and SD}
    \label{fig:2}
\end{figure}
In a within-subject design, each participant serves as their own control, which helps to reduce the impact of individual variables that may affect the results. By using a paired sample t-test, we can analyze the differences between the paired samples and determine observed differences are statistically significant or simply by chance. Upon performing a paired sample t-test with the obtained statistics from data collection and from Figure 3, the results show clear evidence that sample average between with leaderboard and without leaderboard has big difference to be statistically significant.
\begin{figure}[htp]
    \centering
    \includegraphics[width=0.5\textwidth]{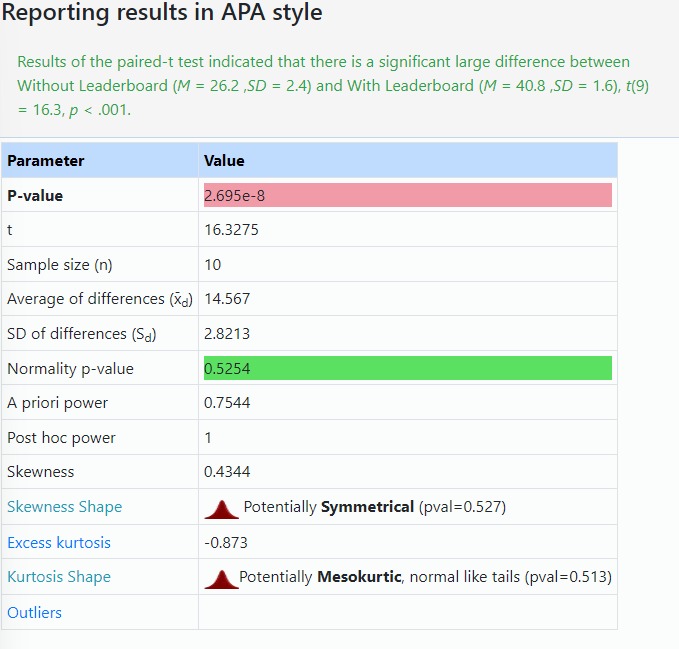}
    \caption{Results of Paired Sample T-Test}
    \label{fig:3}
\end{figure}

\section{Conclusion}

From Figure 2, we can observe that the average user engagement time has increased after implementing the leaderboard to the app by around 68\%. The Figure 3 shows that the t-test proves successfully prove that the difference cannot be by chance but statistically recognizable with leaderboard implementation. This proves that by using such incentivized methods, the web and app developers including educators can improve their existing systems to be more engaging for the users as seen in Figure 4.
\begin{figure}
    \centering
    \includegraphics[width=0.4\textwidth]{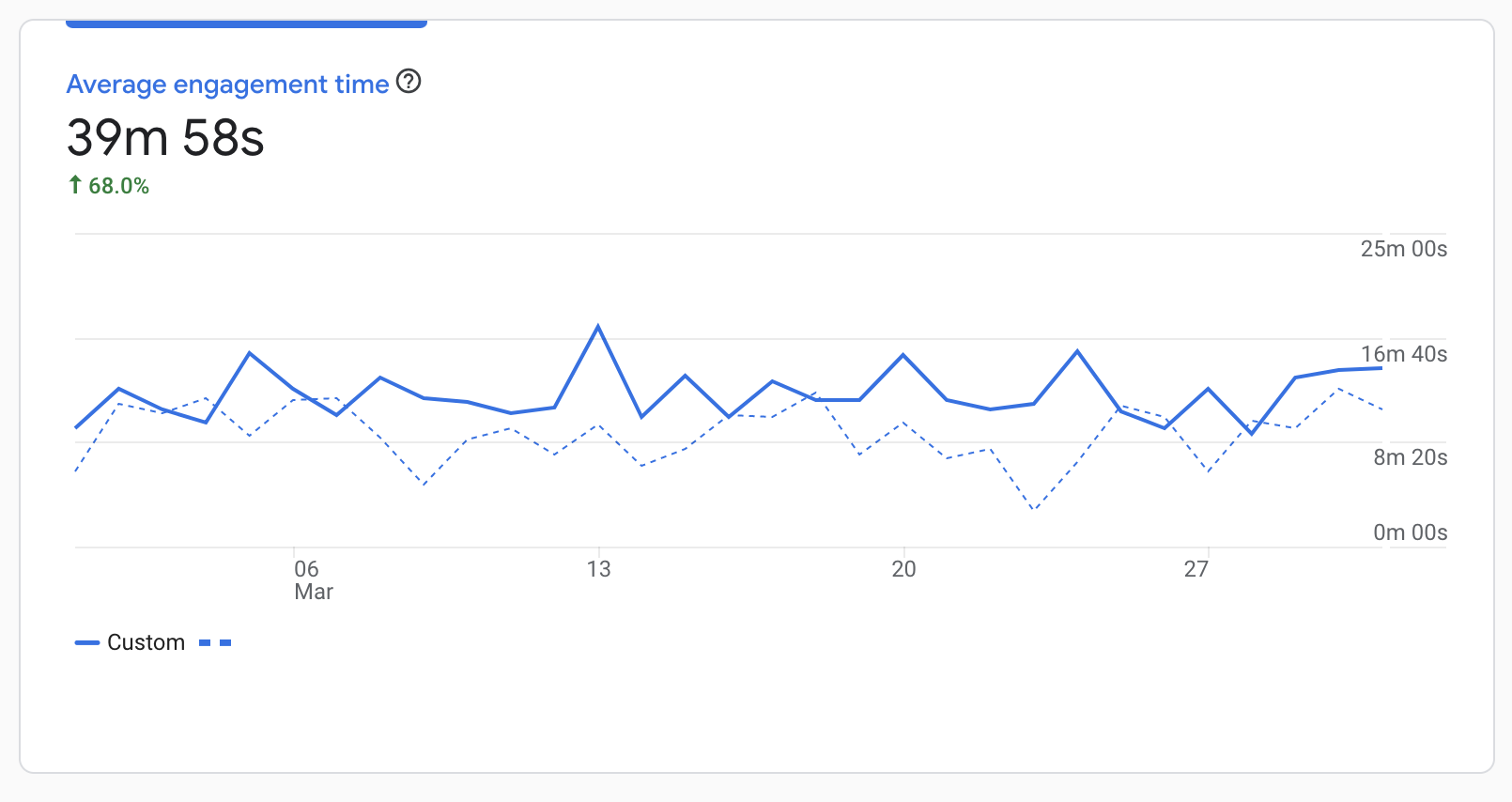}
    \caption{Comparision of Average Engagement Time Without (Dotted) and With (Solid) Leaderboard}
    \label{fig:4}
\end{figure}

The validity of the experiment can be threatened internally if the participant is positioned in last places in the leaderboard that can demotivate them to use the app. Externally, the user engagement automatically increases during the months just before exams which does not have any relation to the leaderboard. In order to make our experiment more valid, we can categorize the questions in such a way that the difficulty level is same irrespective of the level of independent variables to increase the experiment validity.


\bibliography{refs}
\bibliographystyle{plain}

\end{document}